\documentclass{jpsj-suppl}
\usepackage{times}
\usepackage{color}

\title{Nonequilibrium transport through a quantum dot coupled to normal and
superconducting leads}

\author{Akihisa Koga\thanks{E-mail address: koga@phys.titech.ac.jp} 
}
\inst{Department of Physics, Tokyo Institute of Technology, Meguro, Tokyo 152-8551, Japan 
}

\recdate{August 26, 2013}

\abst{
We study the interacting quantum dot 
coupled to the normal and superconducting leads 
by means of a continuous-time quantum Monte Carlo method
in the Keldysh-Nambu formalism.
Deducing the steady current through the quantum dot under a finite voltage,
we examine how the gap magnitude in the superconducting lead and 
the interaction strength at the quantum dot affect transport properties.
It is clarified that the Andreev reflection and Kondo effect lead
to nonmonotonic behavior in the nonequilibrium transport at zero temperature.
}

\kword{Nonequilibrium transport, 
continuous-time quantum Monte Carlo Simulation}

\begin{document}
\maketitle

\section{Introduction}
Electron transport through nanofabrications 
has attracted interests.
One of the interesting systems is the quantum dot system 
coupled to the normal and superconducting leads,
which has experimentally been realized~\cite{ex1,ex2,ex3}.
It has recently been examined how 
the Kondo effect due to electron correlations competes with 
the proximity-induced on-dot pairing effects~\cite{Deacon}.
Theoretical study for the system has been done by many groups
~\cite{slave1,slave2,Domanski,nca,Cuevas,Yamada,Tanaka,Baranski}
and some interesting transport properties have successfully been explained.
However, it is not clear how the current are affected by 
the local correlations and Andreev reflection quantitatively.
This may be crucial to understand the experimental results correctly
since the linear response region, which can be treated quantitatively 
by means of the numerical renormalization group method, is narrow
in the interacting quantum dot system.
Therefore, another unbiased method is desired to discuss 
the nonlinear response in the system. 
One of the appropriate techniques is 
the continuous-time quantum Monte Carlo (CTQMC) 
method~\cite{CTQMC} based on the Keldysh formalism~\cite{WernerOka,Werner}.
In our previous paper~\cite{NONKOGA}, we have used the CTQMC method 
in the Nambu formalism to discuss the nonequilibrium 
transport properties in the quantum dot coupled 
to the normal and superconducting leads.
However, the analysis was restricted in the simple case
and it is still unclear how the superconducting gap 
affects transport properties for the quantum dot system.

To clarify this,
we consider the interacting quantum dot coupled to 
the normal and superconducting leads.
Calculating the time evolution of the current 
through the quantum dot at zero temperature,
we examine how the gap magnitude in the superconducting lead
affects the nonequilibrium transport.
It is clarified that the Andreev reflection and Kondo effect lead
to nonmonotonic behavior in the steady current.

This paper is organized as follows. In Sec. \ref{2}, we introduce
the model Hamiltonian and briefly explain the CTQMC algorithm in the
Keldysh-Nambu formalism. In Sec. \ref{3}, we
discuss the nonequilibrium phenomena in the quantum-dot
system. A summary is given in Sec. \ref{4}.

\section{Model and Method}\label{2}
We consider the interacting quantum dot coupled to 
the normal and superconducting leads.
For simplicity, we use a single level quantum dot 
with the Coulomb interaction $U$ and assume the
superconducting lead to be described by the BCS
theory with an isotropic gap $\Delta$.
The model Hamiltonian should be given as
\begin{eqnarray}
H&=&H_0+H',\label{Hami}\\
H_0&=&\sum_{k\alpha\sigma}\left(\epsilon_{k\alpha}-\mu_\alpha\right)
c_{k\alpha\sigma}^\dag c_{k\alpha\sigma}+
\Delta \sum_k\left(c_{-kS\downarrow}^\dag c_{kS\uparrow}^\dag+
c_{kS\uparrow} c_{-kS\downarrow}\right)\nonumber\\
&+&\sum_{k\alpha\sigma}\left(
V_{k\alpha} c_{k\alpha\sigma}^\dag d_\sigma+
V_{k\alpha}^* d_\sigma^\dag c_{k\alpha\sigma} \right)+
\sum_\sigma\left(\epsilon_d+\frac{U}{2}\right)n_\sigma,\\
H'&=&
U\left( n_\uparrow n_\downarrow 
-\frac{1}{2}\sum_\sigma n_\sigma\right),\label{U}
\end{eqnarray}
where $c_{k\alpha\sigma} (c_{k\alpha\sigma}^\dag)$ 
is the annihilation (creation) 
operator of an electron with wave vector $k$ and 
spin $\sigma(=\uparrow, \downarrow)$ 
in the $\alpha$th lead.
$d_{\sigma} (d_{\sigma}^\dag)$ is 
the annihilation (creation) 
operator of an electron at the quantum dot and 
$n_\sigma=d_\sigma^\dag d_\sigma$.
$\epsilon_{k\alpha}$ is the dispersion relation of the $\alpha$th lead, 
$V_{k\alpha}$ is the hybridization 
between the $\alpha$th lead and the quantum dot, and
$\epsilon_d$ is the energy level.
We set  
the chemical potential in each lead as $\mu_N=V$ and $\mu_S=0$, where
$V$ is the bias voltage.
We here consider the system
with $\epsilon_d+U/2=0$
in the infinite bandwidth limit,
where the coupling 
$\Gamma_\alpha(\omega)=\pi\sum_k\left|V_{k\alpha}\right|^2
\delta(\omega-\epsilon_{k\alpha})$ is constant.

In this study, we use the weak-coupling version of the CTQMC method 
based on the Keldysh formalism~\cite{WernerOka,Werner}.
In the method, we simulate the system prepared 
in the noninteracting nonequilibrium state 
with the interaction turned on at time $t=0$.
Therefore, the simulation may be referred to as an "interaction quench".
When fluctuations due to the interaction quench relaxes and
the system converges,
we can discuss steady-state properties in the framework.

\begin{figure}
\begin{center}
\includegraphics[width=9cm]{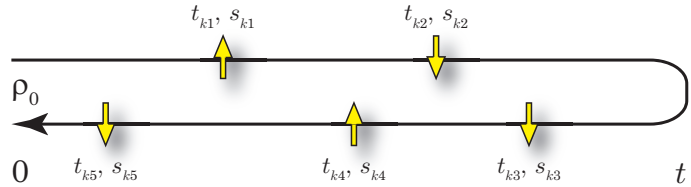}
\end{center}
\caption{(Color online) Illustration of the Keldysh contour for the
CTQMC method. Arrows represent auxiliary Ising spins for a certain
configuration corresponding to the perturbation order
$l = 2$ and $m = 3 \;\;(n = 5)$.}
\label{f}
\end{figure}

We first consider the following identity as
\begin{eqnarray}
1&=&{\rm Tr}\left[\rho_0 e^{it\left(H_0+H'-K/t\right)}
e^{-it\left(H_0+H'-K/t\right)}\right],\label{4}
\end{eqnarray}
where $\rho_0 = e^{-\beta H_0} / {\rm Tr} \Big[e^{-\beta H_0}\Big]$ and
$K$ is a nonzero constant.
By expanding two exponentials in eq. (\ref{4}) in terms of the interaction representation, 
we obtain as
\begin{eqnarray}
1&=&{\rm Tr}\Big\{\rho_0 
\tilde{T}\left[\exp\Big\{i\int_0^t d\tilde{t} \left(H'(\tilde{t})-\frac{K}{t}\right)
\Big\}\right]
e^{itH_0}\nonumber\\
&&\times
e^{-itH_0}
T\left[\exp\Big\{-i\int_0^t dt \left(H'(t)-\frac{K}{t}\right)\Big\}\right]\Big\},\nonumber\\
&=&\sum_l\Big(-\frac{iK}{t}\Big)^l\int_{0}^t d\tilde{t}_1\cdots
\int_{\tilde{t}_{l-1}}^t d\tilde{t}_l
\sum_m \Big(\frac{iK}{t}\Big)^m
\int_{0}^t dt_1\cdots\int_{t_{m-1}}^t dt_m
\nonumber\\
&\times&{\rm Tr}\Big[\rho_0  e^{i\tilde{t}_1H_0}
\Big(1-\frac{t}{K}H'\Big)\cdots e^{i(\tilde{t}_l - \tilde{t}_{l-1}) H_0}
\Big(1-\frac{t}{K}H'\Big) e^{i(t - \tilde{t}_{l}) H_0}\nonumber\\
&\times&
e^{-i(t - t_m)H_0}
\Big(1-\frac{t}{K}H'\Big)\cdots e^{-i(t_2-t_1)H_0}
\Big(1-\frac{t}{K}H'\Big)e^{-it_1H_0}\Big]
\nonumber,
\end{eqnarray}
where $T (\tilde{T})$ is the time-ordering (antitime-ordering) operator.
By using the following equation as
\begin{eqnarray}
1-\frac{t}{K}H'=1-\frac{tU}{K}\left(n_\uparrow n_\downarrow-\frac{1}{2}
\sum_\sigma n_\sigma\right)
=\frac{1}{2}\sum_{s=\pm 1}e^{\gamma s (n_\uparrow-n_\downarrow)},\label{K}
\end{eqnarray}
with $\gamma = \cosh^{-1}(1+tU/2K)$, the identity is represented as
\begin{eqnarray}
1&=&\sum_{lm}(-i)^l i^m\left(\frac{K}{2t}\right)^{l+m}\sum_{\{\tilde{s}\}\{s\}}
\int_0^t d\tilde{t}_1\cdots\int_{\tilde{t}_{l-1}}^t d\tilde{t}_l
\int_0^t dt_1\cdots\int_{t_{m-1}}^t dt_m\nonumber\\
&\times&\!\!\!\!{\rm Tr}\Big[
\rho_0e^{i{\tilde t}_1 H_0}e^{\gamma \tilde{s}_1
(n_\uparrow-n_\downarrow)}\cdots 
e^{\gamma \tilde{s}_l(n_\uparrow-n_\downarrow)}
e^{-i({\tilde t}_l-t_m) H_0}e^{\gamma s_m(n_\uparrow-n_\downarrow)}
\cdots
e^{\gamma s_1(n_\uparrow-n_\downarrow)}
e^{-it_1H_0}\Big].
\end{eqnarray}
The introduction of 
the Ising valuable $s$ in eq. (\ref{K}) allows us to perform Monte Carlo simulations.
An $(l+m)$th order configuration $c=\{s_{k_1}, s_{k_2}, \cdots, s_{k_n};
t_{k_1}, t_{k_2}, \cdots, t_{k_n} \}$ is represented by the auxiliary spins 
$s_{k_1}, s_{k_2}, \cdots, s_{k_n}$ at the Keldysh times 
$t_{k_1}, t_{k_2}, \cdots, t_{k_n}$ along the Keldysh contour,
where the $l (m)$ denotes the number of spins 
on the forward (backward) contour and $n=l+m$ (see Fig. \ref{f}). 
Its weight $w_c$ is then given as
\begin{eqnarray}
w_c &=& (-i)^l i^m\left(\frac{Kdt}{2t}\right)^n 
\det \left[{\hat N}^{(n)}\right]^{-1},\label{weight}
\end{eqnarray}
where ${\hat N}$ is an $n\times n$ matrix
and its element consists of a $2\times 2$ matrix~\cite{KogaWerner} as
$\left[{\hat N}^{(n)}\right]^{-1}={\hat \Gamma}^{(n)}-{\hat g}^{(n)}
\left({\hat \Gamma}^{(n)}-{\hat I}^{(n)}\right)$,
${\hat I}^{(n)}_{ij}=\delta_{ij}{\hat\sigma}_0$, 
${\hat \Gamma}^{(n)}_{ij}=\delta_{ij}
\exp\left({\gamma s_{k_i}{\hat\sigma}_z}\right)$, and
${\hat g}^{(n)}_{ij}={\hat \sigma}_z{\hat G}_0(t_{k_i}, t_{k_j})$,
where ${\hat\sigma}_0$ is the identity matrix and ${\hat\sigma}_z$ is 
the $z$-component of the Pauli matrix.
The matrix ${\hat G}_0$ is given by
the lesser and greater Green's functions as
\begin{equation}
{\hat G}_0(t'_k,t''_k)=\left\{
\begin{array}{cc}
{\hat G}_0^<(t',t'')&t'_k < t''_k\\
{\hat G}_0^>(t',t'')&t'_k\ge t''_k
\end{array}
\right.,
\end{equation}
where 
the times $t'$ and $t''$ correspond to the Keldysh times $t_k'$ and $t_k''$.
These Green's functions have been obtained by 
the standard technique~\cite{Yamada,Cuevas}.
We note that the weight for a certain configuration 
is represented by the complex number.
This should yield serious dynamical sign problem if simulations
are performed on the longer contours.
Therefore, accurate calculations are restricted to a certain time $t_{max}$.

To perform Monte Carlo simulations, 
we use the Metropolis algorithm with the simple sampling process,
where an Ising spin is inserted or removed in the Keldysh contour
in each Monte Carlo step.
Here, we measure the current from the quantum dot to 
$\alpha$th lead $I_\alpha$, which is defined as 
$I_\alpha = -2{\rm Im}\sum_{k\alpha}V_{k\alpha\sigma}
\langle c_{k\alpha\sigma}^\dag d_\sigma\rangle$.
The detail of the measurement formula is given in Ref. \cite{NONKOGA}.
In this study, we use the coupling constant of the normal
lead $\Gamma_N$ as the unit of energy and fix the parameters
as $\Gamma_S/\Gamma_N = 1$, $V/\Gamma_N=0.5$, and $T/\Gamma_N=0$. 
In the following, we perform the
CTQMC simulations to discuss the nonequilibrium transport 
in a quantum dot coupled to normal and
superconducting leads.

\section{Results}\label{3}
\begin{figure}
\begin{center}
\includegraphics[height=5.5cm]{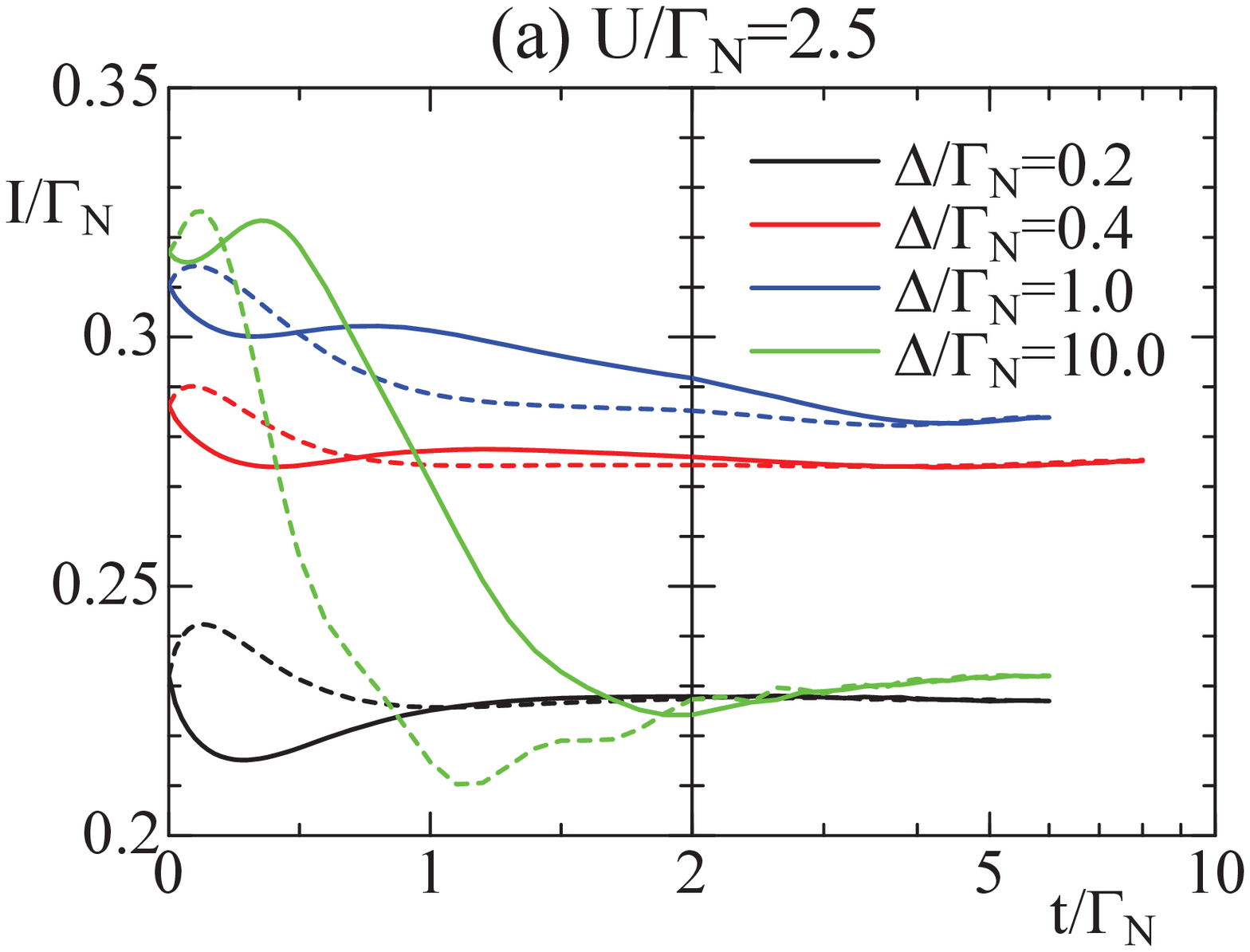}
\includegraphics[height=5.5cm]{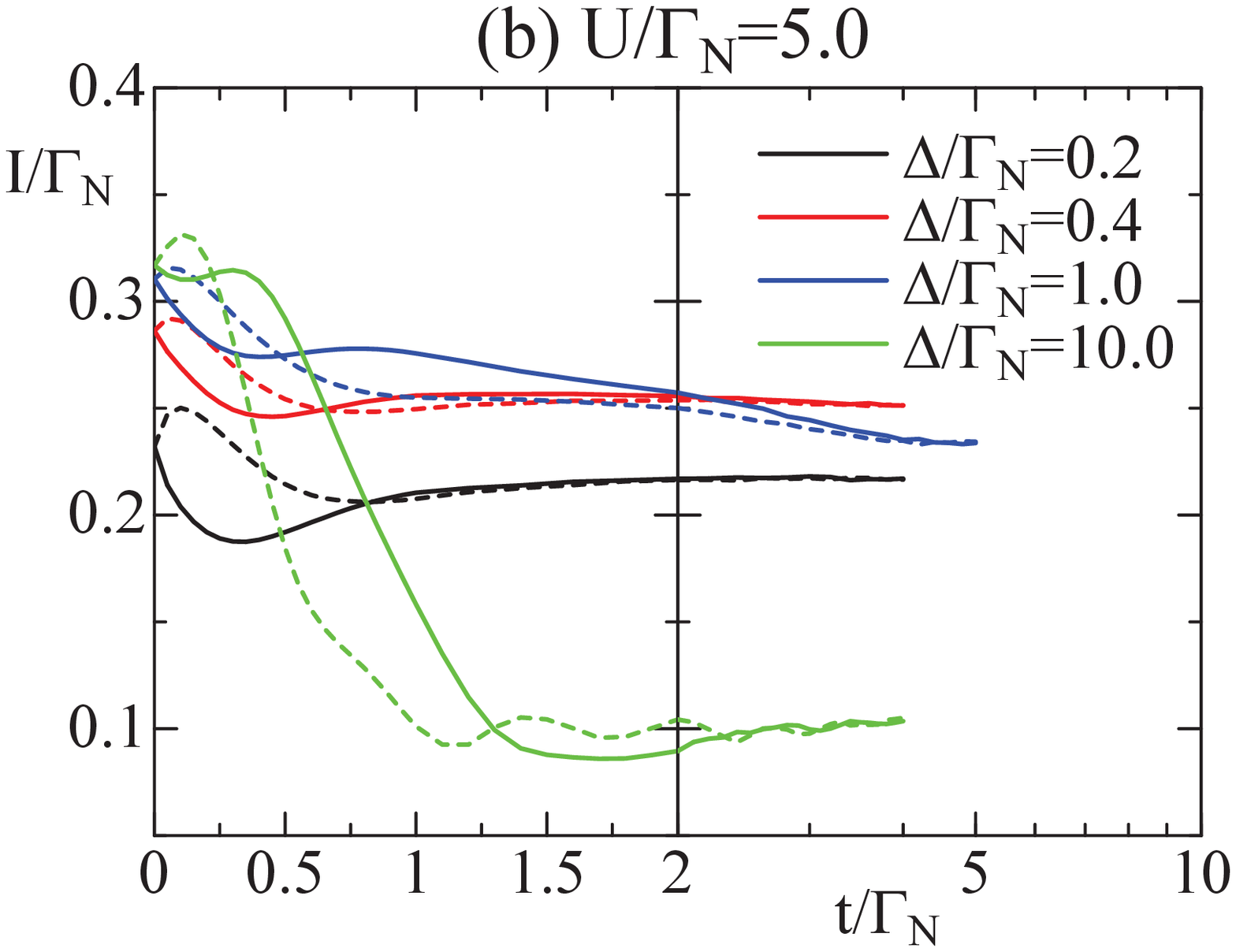}
\end{center}
\caption{(Color online) Time evolution of the currents for the system with 
$V/\Gamma_N=0.5$ and $T/\Gamma_N=0$ when $U/\Gamma_N=2.5 (a)$ and $5.0$ (b).
Solid (dashed) lines represent the currents 
from the quantum dot to the normal (superconducting) lead.
}
\label{f1}
\end{figure}

In the section, we discuss how the gap magnitude affects 
the steady current through the interacting quantum dot.
By performing CTQMC simulations, 
we calculate the time evolution of
the currents $I_N$ and $-I_S$ with a fixed voltage $V/\Gamma_N=0.5$, 
as shown in Fig. \ref{f1}.
In the figure, quantities are shown on the linear plot 
in the initial relaxation region ($t\Gamma_N < 2$) 
and on the logarithmic plot in the rest ($t\Gamma_N > 2$).
When $t\Gamma_N=0$, the steady current $(I=I_N=-I_S)$ flows
through the noninteracting quantum dot.
The introduction of the interaction yields
oscillation behavior in both currents ($I_N$ and $-I_S$).
Although two currents are different in the transient region,
we find that each oscillation behavior is quickly damped and 
these currents approach a certain value when the time proceeds,
as shown in Fig. \ref{f1}.
Therefore, in the case, 
the current at $t=t_{max}$ can be regarded as the steady current.

When $\Delta/\Gamma_N\lesssim 0.5$, 
there is not so large difference between the currents at $t\Gamma_N=0$ and
$t=t_{max}$
although the oscillation behavior appears in the initial relaxation,
as shown in Fig. \ref{f1} (a).
This means that the interaction at the quantum dot little 
affects the steady current.
In fact, the increase of the gap magnitude monotonically increases
the steady current, which is similar to that for the noninteracting case.
Therefore, we can say that when the gap magnitude is small enough, 
the Andreev reflection is dominant and the Kondo effect little affects
transport properties.

On the other hand, when $\Delta/\Gamma_N\gtrsim 0.5$,
the steady current is suppressed by the increase of the gap magnitude,  
as shown in Figs. \ref{f1} (a) and (b).
This behavior may be explained by the following.
The Coulomb interaction at the quantum dot induces 
the Kondo resonance peak around the chemical potential.
On the other hand, the energy level for an electron with an opposite spin
pairing with the electron is away from 
the chemical potential and its density of states decreases due to the Kondo 
effect.
Therefore, the Andreev transport under the finite voltage is strongly 
suppressed.
In a larger gap case, the Andreev current little flows and 
the system may be regarded as an insulating state.

\begin{figure}
\begin{center}
\includegraphics[width=12cm]{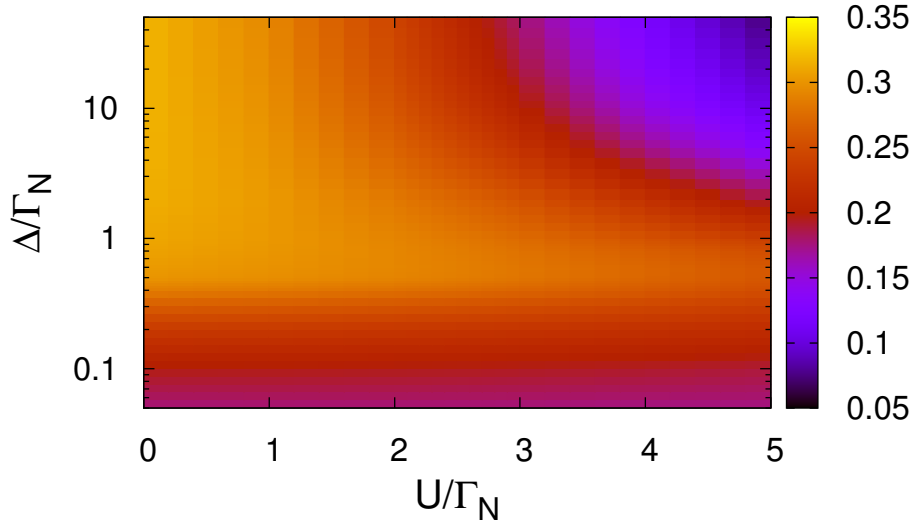}
\end{center}
\caption{(Color online) Density plot of the normalized steady current 
$I/\Gamma_N$
through the quantum dot system with $V/\Gamma_N=0.5$ and $T/\Gamma_N=0$.}
\label{f2}
\end{figure}

By performing similar calculations, 
we obtain the density plot of the steady current, as shown in Fig. \ref{f2}.
In the noninteracting case with $U/\Gamma_N=0$, 
the increase of the gap magnitude monotonically increases the steady current,
which should be induced by the Andreev reflections.
On the other hand, the introduction of the interaction leads 
to different behavior, where the development of the Kondo peak 
around the chemical potential suppresses the Andreev reflection.
Therefore, in the larger $U$ and $\Delta$ region, 
the steady current is strongly suppressed.

\section{Summary}\label{4}
We study nonequilibrium transport through the interacting quantum dot 
coupled to the normal and superconducting leads 
by means of a continuous-time quantum Monte Carlo method
in the Keldysh-Nambu formalism.
Calculating the time evolution of the current through the quantum dot,
we discuss how the gap magnitude in the superconducting lead 
and the interaction at the quantum dot affect the steady current.
We have found that nonmonotonic behavior is induced by the competition
between the Kondo effect and the Andreev reflections.

\section*{Acknowledgments}
This work was partly supported by Japan Society for the Promotion of Science 
Grants-in-Aid for Scientific Research Grant Number 25800193
and the Global COE Program ``Nanoscience and Quantum Physics" from 
the Ministry of Education, Culture, Sports, Science and Technology (MEXT) 
of Japan. 
A part of computations was carried
out on TSUBAME2.0 at Global Scientific Information
and Computing Center of Tokyo Institute of Technology
and on the Supercomputer Center at the Institute
for Solid State Physics, University of Tokyo. The simulations
have been performed using some of the ALPS libraries~\cite{alps}.

\end{document}